\documentclass[prl,twocolumn]{revtex4}
\usepackage{graphicx, epsfig}
\usepackage{color}
\usepackage{mathrsfs}

\newcommand{\be}{\begin{equation}}
\newcommand{\ee}{\end{equation}}
\newcommand{\bea}{\begin{eqnarray}}
\newcommand{\eea}{\end{eqnarray}}
\newcommand{\ba}{\begin{eqnarray}}
\newcommand{\ea}{\end{eqnarray}}

\newcommand{\gapp}{\mathrel{\raise.3ex\hbox{$>$}\mkern-14mu
              \lower0.6ex\hbox{$\sim$}}}
\newcommand{\lapp}{\mathrel{\raise.3ex\hbox{$<$}\mkern-14mu
              \lower0.6ex\hbox{$\sim$}}}

\begin{document}
\title{Dark Strings}

\author{Tanmay Vachaspati}
\affiliation{
CERCA, Department of Physics, 
Case Western Reserve University, Cleveland, OH~~44106-7079.\\
Institute for Advanced Study, Princeton, NJ 08540. 
}

\begin{abstract}
\noindent
Recent astrophysical observations have motivated novel
theoretical models of the dark matter sector. A class 
of such models predicts the existence of GeV scale cosmic 
strings that communicate with the standard model sector 
by Aharonov-Bohm interactions with electrically charged 
particles. We discuss the cosmology of these ``dark strings''
and investigate possible observational signatures.
More elaborate dark sector models are argued to contain 
hybrid topological defects that may also have observational 
signatures.
\end{abstract}

\maketitle

Recent observations by the PAMELA \cite{Adriani:2008zr} and 
ATIC \cite{:2008zzr} experiments show anomalous positron fractions 
and total electron and positron ($e^+e^-$) fluxes in cosmic rays 
in the GeV to TeV energy range. The signal may be an outcome of 
annihilating dark matter \cite{dmlit,ArkaniHamed:2008qn}
or have an astrophysical 
origin 
\cite{Profumo:2008ms}, and both possibilities are 
currently under intense investigation. If the signal is due to 
dark matter annihilation, model building suggests a dark sector 
separate from the standard model sector with a tiny interaction 
linking the two sectors \cite{ArkaniHamed:2008qn}. Thus the
Lagrangian has the form
\begin{equation}
L = L_{\rm SM} + L_{\rm DS} + L_{\rm int}
\label{Lag}
\end{equation}
where SM stands for standard model, DS for dark sector, and
``int'' represents the interaction that bridges between the
two sectors. The dark sector has its own symmetries that are 
{\it all} spontaneously broken. This has to be so for Abelian
symmetries, otherwise we would also have a cosmological 
background of massless dark photons that, in the simplest
scenarios, interfere with big bang nucleosynthesis. The issue 
is more subtle for
non-Abelian symmetries since these can be confining. If the
confinement scale is high enough, the symmetry need not be
broken and dark matter may also be made of ``dark hadrons''.
However, as we shall see, to connect with the standard model
at very low energies, a definite gauge field needs to be picked 
out and the low energy symmetry of the dark sector is effectively
Abelian and broken.

To connect with observations, one of the dark sector symmetries 
is broken at the GeV scale, with a gauge boson that acquires 
mass at the GeV scale. Let us call this gauge boson, $a_\mu$,
and denote its field strength by, $a_{\mu \nu}$. Then the 
interaction Lagrangian is taken to be
\begin{equation}
L_{\rm int} = +\frac{\epsilon}{2} a_{\mu\nu} A^{\mu\nu}
\label{Lint}
\end{equation}
where $A_\mu$ denotes the electromagnetic gauge field and
$A_{\mu\nu}$ its field strength. The coupling $\epsilon$ is
small and is needed to be on the order of $10^{-3}$ 
to connect with observations \cite{ArkaniHamed:2008qn}. 
We shall take $\epsilon = 10^{-3} \epsilon_3$.

Although more complicated versions of the interaction 
Lagrangian are possible, especially in the context of large 
symmetries and representations in the dark matter sector, they 
all reduce to the form in Eq.~(\ref{Lint}) at energies below 
the GeV scale. The low energy dark sector Lagrangian 
($L_{\rm DS}$) is therefore an Abelian-Higgs model for the 
single relevant gauge field $a_\mu$. The $U(1)_{\rm DS}$ 
symmetry of $L_{\rm DS}$ is broken at the GeV scale and 
$a_\mu$ then acquires a GeV scale mass. The $U(1)_{\rm DS}$ 
symmetry breaking also produces cosmic strings 
\cite{Nielsen:1973cs} with GeV scale tension: 
$\mu \sim (1 ~ {\rm GeV})^2 \sim 10^{-10}{\rm gm/cm}$. The 
only way these strings can interact with the standard model sector 
is via the interaction in $L_{\rm int}$. (Hidden-sector
topological defects have also been considered in 
\cite{Berezinsky:1999az}.)

It is convenient to rewrite the Lagrangian for the gauge
sector in the following manner
\begin{equation}
L_{\rm gauge} 
  = -\frac{1}{4} {\bar A}_{\mu \nu} {\bar A}^{\mu \nu}
    -\frac{1}{4} (1-\epsilon^2) a_{\mu \nu} a^{\mu \nu} 
\end{equation}
where we have absorbed the interaction term into a re-defined
gauge field ${\bar A}_\mu = A_\mu - \epsilon a_\mu$.
This means that a particle with electromagnetic charge
$q$ interacts with $a_\mu$ with an effective dark sector 
charge of $\epsilon q$.

A dark string with unit topological winding contains $2\pi / e'$ 
quanta of magnetic flux where $e'$ is the dark sector unit charge. 
Hence the Aharonov-Bohm (AB) phase around a unit winding string is
$\phi_{\rm AB} = {2\pi \epsilon q}/{e'}$.
If $\epsilon q$ is an integer multiple of $e'$, there will be
no AB interaction. However, there is no reason for such a relation 
to hold, since the value of $\epsilon$ is set by integrating out 
heavy degrees of freedom. To be specific, we shall assume $q = e'$ 
and hence that the AB phase is $2\pi \epsilon$.

A dark string moving through a medium will encounter friction
due to particle scattering off the string core \cite{Everett:1981nj}
as well as due to AB scattering \cite{Alford:1988sj}. 
The transport cross-section i.e. the cross-section 
that determines momentum transfer, due to AB scattering is
(see \cite{VilenkinShellard})
\begin{equation}
\sigma_{\rm t,AB} = \frac{2}{p} \sin^2(\pi \epsilon) \ ,
\end{equation}
while the transport cross-section due to conventional scattering 
of particles interacting with the string core is
\begin{equation}
\sigma_{\rm t,con} = \frac{\pi^2}{p [\ln (p\delta)]^2} \ ,
\end{equation}
where $\delta \sim 1/\sqrt{\mu}$ is the thickness of the string 
and $p$ is the magnitude of the momentum of the incoming particle.
Hence the friction exerted by AB scattering is larger than the 
conventional drag by the factor
\begin{equation}
\frac{F_{\rm AB}}{F_{\rm con}} = 
\frac{2}{\pi^2} \sin^2(\pi \epsilon) \ln^2 (p\delta) \times
\frac{n_{\rm elec}}{n_{\rm DM}}
\end{equation}
where we have also included the last factor that accounts
for the different number densities of electrically charged
particles to dark matter.
The cosmic ratio of baryon to dark matter energy density is 
$\sim 1/6$ while the mass of the dark matter particle 
to nucleon mass is $\sim 10^3$. Putting these factors 
together $ F_{\rm AB}/F_{\rm con} \sim \epsilon_3^2$,
where we have taken $\ln^2(q \delta) \sim 10^3$. So the two 
drag forces are similar in magnitude if $\epsilon_3 \sim 1$.

In the cosmological setting, dark strings form when the cosmic 
temperature drops to $\sim {\rm GeV}$, at time 
$t_f \sim 10^{-5}$ s. The string 
network contains a distribution of loops and infinite strings 
\cite{Vachaspati:1984dz}. In the period following formation, 
the ambient cosmological medium scatters off the strings and 
damps their motion. The frictional damping force (per unit 
length) is $F_d \sim \epsilon^2 n \gamma v$ where $n$ is the 
number density of (electrically charged) particles, $v$ is the 
string speed, and $\gamma$ the Lorentz factor. The rate of work 
done by the damping force is $\sim F_d v$ and this is also the 
rate at which the string loses kinetic energy $\sim \mu v^2$. 
Therefore the frictional damping time scale is 
\begin{equation}
\tau \sim \frac{\mu}{\epsilon^2 n_{\rm elec}} 
\label{dampingtime}
\end{equation}
The number density of particles, $n_{\rm elec}$ is found by 
taking the present number density in protons (and electrons)
and evolving it back in time
\begin{equation}
n_{\rm elec}(t) \sim \frac{\rho_b (t_0)}{m_n}
                  \biggl ( \frac{a_0}{a_{(t)}} \biggr )^3
             \sim 10^{-6} \biggl ( \frac{T}{T_0} \biggr )^3
                                {\rm cm}^{-3} \ ,
\label{nelec}
\end{equation}
where we have used that the density in baryons is $\sim 5\%$ 
of the critical density ($10^{-29} {\rm gms/cm}^3$), the
mass of the nucleon is $m_n \sim 10^{-24} {\rm gms}$,
and the fact that cosmic temperature scales inversely
as the scale factor. Inserting the expression for 
$n_{\rm elec}$ in Eq.~(\ref{dampingtime}),
with $\mu = (1 {\rm GeV})^2$, we find that the damping 
time coincides with the Hubble time at $t_d \sim 1$ s
i.e. at a temperature $\sim 1$ MeV. Hence the strings are 
friction dominated until $t_d$ and subsequently, friction
damping drops rapidly (as $T^3$) and can be neglected. 
Note that the MeV scale coincides with the time at which 
the electrons are becoming non-relativistic and so we are 
justified in using the non-relativistic number density in 
Eq.~(\ref{nelec}).

In the friction dominated regime, string loops smaller than
the horizon will dissipate their energy into heat and collapse.
Infinite strings will smooth out on the damping length scale
$\sim  t_d$. For $t > t_d$, frictional damping becomes unimportant 
and we can ignore it. (The evolution of long strings with friction
has been considered in \cite{Martins:1995tg}.) In the undamped
regime, it is generally believed that strings evolve into a 
network that has ``universal'' scaling properties and 
minimal radiation into massive modes and we will also proceed 
under this assumption. (See, however, 
Ref.\cite{Hindmarsh:2008dw}.)

Loops or dark string can dissipate their energy into
gravitational radiation, $e^+e^-$ pairs, or photons.
The power emitted in gravitational radiation is 
$\sim \Gamma G\mu^2$ where $\Gamma \sim 100$. Since 
$G\mu \sim 10^{-38}$ for us, loops of length 
$l > l_{\rm grav} = 10^{-9} {\rm cm}$ survive longer than 
the current age of the universe ($\sim 10^{17}$ s).
In addition to gravitational radiation, we can expect string 
loops to emit charged particles due to the AB interaction 
\cite{Alford:1988sj}. For a massless charged particle,
dimensional arguments imply a rate of energy loss
$\sim \epsilon^2/l^2$ since the length of the loop is the
only dimensional scale in the problem. (The string width and
tension are not expected to play a role in the AB
interaction.) However, we expect AB 
radiation to be exponentially suppressed due to the mass of 
the electron and only be important for loops with length less 
than the Compton wavelength of the electron, 
$m_e^{-1} \sim 10^{-11}$ cm. 
A higher order interaction, where the string produces
a virtual $e^+e^-$ pair that then annihilate to produce
a photon, is not exponentially suppressed \footnote{``Dark
strings ain't dark'' due to quantum AB radiation.}. The 
power emitted is estimated as
\begin{equation}
\frac{dE_\gamma}{dt} \sim \frac{\epsilon^2}{l^2}
                         \frac{\alpha}{(m_e l)^4} 
\end{equation}
where the first factor is the AB production of massless charged 
particles, the second factor is due to the electron propagators 
in the fermion loop, and the fine structure constant is due to 
the photon vertex. The power radiated in photons falls off rapidly 
with loop length and we find that loops larger than 
$l_\gamma \sim 10^{-8}$ cm will survive longer than the age of 
the universe.

At $t\sim 1 s$, the smallest loop surviving the friction
dominated epoch has size $\sim 10^{10}$ cm. Radiation is
insignificant for such loops.
So we conclude that all (non-self-intersecting) loops from 
$t \sim 1$ s will survive until the present epoch. In the 
meantime, the loops will come to rest in the fluid frame because 
Hubble expansion will damp out any peculiar velocities that the 
loops may have had initially.  This means that loops will 
participate in structure formation, just like any other dark 
matter candidate. They will accrete into galaxies and 
get embedded into stars and other astronomical bodies. 

During the friction-dominated epoch, string curvature on
scales below $t_d$ is damped out, erasing loops with size
less than $t_d$. At $t=t_d$ we expect only
straight strings on the horizon scale and we
can parametrize the number of straight strings per horizon
by $N_s$. Simulations of the string network suggest 
$N_s \sim 10$ in the absence of friction. With friction
included, it seems reasonable that $N_s$ should depend
on the string density at formation since strings move more 
slowly and interact less with each other. We will treat 
$N_s$ as a free parameter. 

The ratio of loop number density to baryon number density at 
$t_d$ is $n_l (t_d)/n_b (t_d) \sim 10^{-54} N_s$ where 
we have taken the loop density to be $\sim N_s/t_d^3$.
Since the loop number density scales like dust, the number of 
loops in the Milky Way ($\sim 10^{12}$ solar masses 
or $10^{45}$ gms) is the number of protons in the Milky Way
($10^{69}$) times $10^{-54}N_s$. From here we estimate that 
gravitational accretion provides the Milky Way with $\sim 10^{15}N_s$ 
loops of size $\sim 10^{10}$ cm and mass $\sim 1$ gm. 

Now we turn to observational signatures of dark strings. The
scenario is somewhat different from that of superconducting 
strings \cite{Chudnovsky:1988cv} because
those strings are frozen into the galactic plasma, while dark 
strings merely experience friction due to the plasma. However,
inhomogeneities within the galaxy can perturb the dynamics of 
a loop and several scenarios are possible as we now discuss.

(i) From Eq.~(\ref{dampingtime}) we find that the damping time
in the galaxy (density $10^{-24}~ {\rm gms/cm^3}$) is $10^{24}$ 
s i.e. longer than the age of the universe. However, random 
perturbations due to density inhomogeneities in the Milky Way 
may cause loops to fragment into many smaller loops which 
are small enough to be insensitive to the medium. These 
loops then survive, occasionally collapsing to give off a
burst of $e^+e^-$. A loop would be very hard to detect 
unless there is a close encounter. Loops that enter
the Earth's atmosphere (density $10^{-3}~ {\rm gms/cm^3}$) have
a dissipation time $10^3$ s and propagate right through the 
atmosphere within this time if they are moving with velocities 
typical in the galaxy, $v \sim 10^{-3} = 10^2 ~{\rm km/s}$. 
On arriving at the Earth's surface (ocean/ground/ice), the loop 
encounters a larger density $\sim 1 ~ {\rm gm/cm^3}$ and the 
dissipation time is correspondingly smaller $\sim 1$ s. In this 
time, a loop will propagate $\sim 100$ km and dissipate its 
energy. The average energy deposited in the
track through the Earth's surface is the loop energy divided 
by the path length i.e. $\sim 10^{-4}~ {\rm ergs/cm} \sim 10^{-2} ~ 
{\rm eV/cm}$ for a loop of length $10^{-8}$ cm. Along the way, 
the loop can also (partially) collapse and produce $\sim 1$ GeV 
positrons that would then annihilate to give gamma rays in the 
ocean or ice. 

To calculate the flux of loops in this scenario, we assume that
the number density of loops increases with decreasing length
until $l \sim l_\gamma$, below which the number density decreases
because the loops are evaporating. Hence loops with 
$l \sim l_\gamma$ are the most numerous and may be expected to
contain a significant fraction of the total string length in
the galaxy. A simple estimate of the number density of loops 
is therefore obtained by considering all the $10^{25}N_s$ cm of 
available string length to be in loops of $\sim 10^{-8}$ cm. Taking 
the loop velocity to be $\sim 10^{-3}$, this implies a galactic 
loop number density $\sim 10^{-33}N_s ~ {\rm cm}^{-3}$ and a flux 
$\sim 10^{-9}N_s ~{\rm km}^{-2} {\rm yr}^{-1}$, which is too small 
to be of experimental interest unless $N_s \gtrsim 10^9$. 
In this case, it would be worth exploring if cosmic ray experiments 
located on the polar ice caps or in the oceans may be sensitive to 
dark string loops.

(ii) The second possibility is that when loops enter dense 
regions of the galaxy, density perturbations cause them to collapse 
and annihilate. One situation in which a loop collapses is known:
if any loop starts at rest and obeys the Nambu-Goto 
equations, then it must collapse into a double line at one instant 
\cite{Kibble:1982cb}. So if some part of a loop enters
a dense region and slows down due to damping, a (partial) collapse 
seems likely. Loop annihilation will produce dark sector 
bosons that will eventually decay into $e^+e^-$. Assuming that the 
$e^+e^-$ are produced with energy $\sim 1$ GeV (i.e.  $10^{-24}$ gms), 
a 1 gm loop injects on order $10^{24}$ $e^+e^-$.
For loops of size $\sim 1$ s, the burst lasts for $\sim 1$ s, 
with an energy output $\sim 10^{21}$ ergs. 
Once injected into the galactic medium, the $e^+e^-$ evolution
is as for secondary $e^+e^-$ production by cosmic rays. 

Even if the energy output of any individual loop collapse
in the galaxy is too small to be of any consequence, 
the cumulative effect of all loops gives rise to a population of 
positrons. The maximum number density of positrons 
produced in this way is estimated by converting the total energy in 
loops in the galaxy, $\sim 10^{15} N_s$ gms, in terms of $\sim 1$ GeV 
positrons. Therefore there are $10^{39}N_s$ positrons and the average 
number density is $10^{-27} N_s~{\rm cm}^{-3}$ where we have taken 
the volume of the galaxy to be $\sim (10 ~ {\rm kpc})^3$. This
number density corresponds to a flux 
$\sim 10^{-17} N_s ~{\rm cm}^{-2} {\rm s}^{-1}$ and is tiny 
compared to the observed flux (though at 10 GeV), which is
$\sim 10^{-5} ~{\rm cm}^{-2} {\rm s}^{-1} {\rm GeV}^{-1}$
\cite{:2008zzr}. 

We could also consider inter-galactic loops that enter the Milky 
Way  -- just like cosmic rays hit the Earth's atmosphere. If the 
structures in the Milky Way, or the dark matter in its halo, 
cause the entering loop to collapse, they would produce a burst 
of $e^+e^-$, and subsequently synchrotron radiation and gamma rays. 
This may result in excess synchrotron radiation from the outskirts 
of the Milky Way and correlated gamma ray emission. 

(iii) The third possibility is that the loop dynamics remains 
largely unaffected in the galactic medium and loops simply reside
and oscillate in the galaxy. This possibility implies that there 
is a population of loops in our Milky Way but it is not clear how 
one might observe them. The only hope seems to be if loops
develop cusps -- points on the string that reach the speed of
light -- that beam large amounts of electromagnetic radiation 
via the AB interaction. This may be similar to radio transients
(``sparks'') from superconducting strings \cite{Vachaspati:2008su}. 
Cusp annihilation would also produce a beam of energetic $e^+e^-$ 
that might perhaps be observable, similar to the discussion in 
Ref.~\cite{Brandenberger:2009ia} but adapted for GeV scale strings 
within our galaxy.

In the string evolution scenario of 
Refs.~\cite{Hindmarsh:2008dw},
the string network maintains scaling by releasing all its energy 
directly to particles. In our case, this would imply that the string 
network loses energy by directly emitting dark sector particles 
that eventually convert to $e^+e^-$, and little radiation in 
electromagnetic and gravitational radiation. 
The $e^+e^-$ inject energy into the cosmological medium and
one might hope for an observational signature on this account. 
At the epoch of recombination, the $e^+e^-$ may cause enhanced
reionization and this would affect the propagation of cosmic 
microwave background photons \cite{Zhang:2007zzh}. 
If we evolve the loops at $t=t_d$ like pressureless matter, 
the energy density in GeV scale cosmic strings 
at recombination is only $10^{-30} N_s$ of the baryon energy 
density. Even if all the string energy gets converted to ionizing
radiation, this amounts to $\sim 10^{-20} N_s$ eV per atom
in the cosmological medium. Hence enough energy is injected 
to reionize 1 in $10^{21} N_s^{-1}$ atoms and this is miniscule 
compared to say the number density of $e^-$ due to residual 
ionization at recombination ($10^{-5}$).

Let us now discuss non-minimal models of the type in
Eq.~(\ref{Lag}). More generally the dark sector 
has a symmetry group, $G_{\rm dark}$, that
is completely broken at low energy scales. We have 
considered the case $G_{\rm dark} = U(1)$ so far but
$G_{\rm dark}$ could be larger. In this case, the dark
gauge field in $L_{\rm int}$, $a_\mu$, is one of a gauge 
multiplet and has to be selected e.g.  $\Phi^a a_\mu^a$ where 
the index $a$ is the group index and the adjoint field
$\Phi^a$ gets a vacuum expectation value (VEV). However,
the VEV of an adjoint scalar field leads to an unbroken
$U(1)$ and hence ``dark magnetic monopoles''. 
When the surviving $U(1)$ is broken at
the GeV scale, the $U(1)$ magnetic flux is confined
and the monopoles get connected by strings. 
The string segments, with monopoles at either end,
annihilate to produce $e^+e^-$ with energy at the scale 
of the monopoles i.e. VEV of
$\Phi^a$ which, for concreteness, we take to be the TeV
scale. Eventually the annihilation of positrons gives
TeV gamma rays. If these gamma rays are produced relatively
late (close to recombination), they may not thermalize and
would be constrained by the observed gamma ray background,
as discussed in \cite{Brandenberger:2002qh}.

In addition to the string segments that connect monopoles, 
closed loops of strings will also be produced and the length 
distribution of loops depends on the number density of 
monopoles that are present during string formation. In 
particular, if the monopoles
have been inflated away, the string distribution is
identical to the $U(1)$ case and the closed loops of strings
will follow the evolution discussed above in the context
of the $U(1)$ model. On the other hand, if the monopoles are
relatively dense during string formation, very few loops are
formed and the segments connect nearby monopoles and are 
short. (Strings can break by nucleating monopole-antimonopole 
pairs but this process is suppressed by the factor 
$\exp(- \pi m^2/\mu )$ \cite{Vilenkin:1982hm} and can be 
neglected for modest separation of monopole and string energy 
scales.)

If GeV scale dark strings exist, it may also be possible to 
create them in the laboratory via the AB interaction by 
scattering electrically charged particles. A $10^{-11}$ cm loop 
has $\sim 1$ TeV energy, well within the energy range of existing
colliders. However, studies of kink creation in 1+1 dimension
\cite{Dutta:2008jt} suggest that low energy scattering of a large 
number of electrically charged particles, such as atomic nuclei, 
may be more suitable for creating solitons than two particle
scatterings.

Finally, even if the observed positron signatures have an 
astrophysical origin, our discussion applies quite generally to 
models where there is a separate dark sector of the kind given in 
Eq.~(\ref{Lag}). Dark strings may provide a novel window to this 
class of models.

I am grateful to Peter Graham for discussions that led to 
this work and for very helpful feedback. I thank 
Nima Arkani-Hamed, Juan Maldacena, Harsh Mathur, 
Scott Thomas, Alex Vilenkin and Edward Witten for comments.
This work was supported by the U.S. Department of Energy at 
Case Western Reserve University, and grant number 
DE-FG02-90ER40542 at the Institute for Advanced Study.

\end{document}